\documentclass{amsart} 

\usepackage{amssymb, amsmath, sidecap}
\usepackage{amsfonts}
\usepackage{float}
\usepackage{graphics}
\usepackage{graphicx}
\usepackage{subfigure}
\usepackage{verbatim}
\usepackage{textcomp}
\usepackage{color}
\usepackage{caption2}

\def\bt{\begin{thm}}
\def\et{\end{thm}}
\def\bl{\begin{lem}}
\def\el{\end{lem}}
\def\bd{\begin{defi}}
\def\ed{\end{defi}}
\def\bc{\begin{cor}}
\def\ec{\end{cor}}
\def\bp{\begin{proof}}
\def\ep{\end{proof}}
\def\br{\begin{rem}}
\def\er{\end{rem}}

\newtheorem{thm}{Theorem}[section]
\newtheorem{lem}{Lemma}[section]
\newtheorem{defi}{Definition}[section]

\newtheorem{rem}{Remark}[section]
\newtheorem{cor}{Corollary}[section]

\newcommand{\pd}[2]{\frac{\textstyle \partial #1}{\textstyle \partial #2}}

\newcommand{\fraction}[2]{\frac{\textstyle #1}{\textstyle #2}}

\numberwithin{equation}{section}

\graphicspath{{./figures/}}

\title[Cahn-Hilliard Model with Long-Range Repulsive Interactions]{Dynamic Transitions and Pattern Formations for Cahn-Hilliard Model with Long-Range Repulsive Interactions}
\author[Liu]{Honghu Liu}
\address[Liu]{Department of Mathematics,
Indiana University, Bloomington, IN 47405}
\email{liu40@umail.iu.edu}

\author[Sengul]{Taylan Sengul}
\address[Sengul]{Department of Mathematics,
Indiana University, Bloomington, IN 47405}
\email{msengul@indiana.edu}

\author[Wang]{Shouhong Wang}
\address[Wang]{Department of Mathematics,
Indiana University, Bloomington, IN 47405}
\email{showang@indiana.edu, http://www.indiana.edu/\texttildelow fluid}

\author[Zhang]{Pingwen Zhang}
\address[Zhang]{Department of Mathematics, Beijing University,
Beijing, P. R. China}
\email{pzhang@pku.edu.cn}

\thanks{The work  was supported in part by the
Office of Naval Research and by the National Science Foundation.}

\keywords{Cahn-Hilliard model with long-range repulsive interactions,  diblock copolymer melts, dynamic transition theory, pattern formation, hexagonal patterns, metastability, phase diagrams}
\subjclass{}

\begin{document}

\begin{abstract}
The main objective of this article is to study the order-disorder phase transition
and pattern formation for systems with long-range repulsive interactions. The main focus is on the  Cahn-Hilliard model  with a nonlocal term in the corresponding energy functional, representing the long-range repulsive interaction.  First, we show that as soon as the linear problem loses stability,
the system always undergoes a dynamic transition to one of the three
types,  forming different patterns/structures. The types of
transition are then dictated by a nondimensional parameter,
measuring the interactions between the long-range repulsive term and
the quadratic and cubic nonlinearities in the model. The derived explicit form of this parameter offers precise information for the phase diagrams. 
Second, we obtain a novel and explicit pattern selection
mechanism associated with the competition between the long-range
repulsive interaction and the short-range attractive interactions. In particular, the hexagonal pattern is unique to the long-range interaction, and is associated with a novel 
two-dimensional reduced  transition equations on the center manifold generated by the unstable modes, consisting of (degenerate) quadratic terms and
non-degenerate cubic terms. Finally, explicit information on the metastability and basin of attraction of different disordered/ordered states and patterns are derived as well. 
\end{abstract}

\maketitle
\tableofcontents

\section{Introduction}
Many systems in nature can be modeled through the inclusion of
long-range interactions, examples including uniaxial
ferromagnetic films, Langmuir monolayers, block copolymers, and
cholesteric liquid crystals. New features appear in the phase
transition in such systems where long-range repulsive interactions
compete with the short-range attractive interactions. Competing
interactions can lead to the emergence of modulated phases, where a
particular pattern and wavelength are selected \cite{DK09, SA95}.
In this paper, we will concentrate on the effects of
such an interaction on the phase transition in the diblock copolymer melts.

A diblock copolymer is a linear chain macromolecule composed of two
subchains of chemically distinct repeat monomers joined covalently
to each other. One of the subchains is formed of type-A monomers,
and the other of type-B monomers. Diblock copolymer melts have been
studied for decades by polymer chemists due in large part to their
ability to assemble into various ordered structures at low
temperatures. However, unlike many other binary systems, e.g.
polymer blends or polymer solutions, in which macroscopic transition
occurs, the phase transition here is on a mesoscopic scale where the
microdomains of A-rich and B-rich regions emerge, and the domains
observed have highly regular periodic structures such as lamellae
(LAM), hexagonally packed cylinders (HPC), spheres, double
bicontinuous gyroid (DG), and Fddd patterns \cite{BF99, BM96, JWHZ}.

Through decades of experimental and theoretical investigations, a
number of elegant theories have been developed to describe the
behavior of diblock copolymer melts. Most of them are based on mean
field approximation; see among many others \cite{BF90,BF99, BM96,
deG79,FVD02, JWHZ, Kaw04, OE97}. Also, the classical phase diagram
predicted by mean field theories or constructed by experiments
indicates several regimes for the phase separation; see  \cite{MS94,
FVD02, KFBHRBAM95} and references therein.

The main objective of this article is to study the phase transition
and pattern formation associated with the order-disorder transition
(ODT) of noncrystalline diblock copolymer melts {and to analyze
the effects of the long-range repulsive interactions on the phase
transitions}.

The modeling is based on the following considerations. First, for simplicity, we neglect
the polydispersity effect, i.e., each polymer in the melt has the
same polymerization number $N$ and the same number of type-A
monomers. Also, we assume  that both types of monomers share the
same Kuhn statistical length $l$, which measures the statistical
distance between two adjacent monomers in the same subchain. The model we adopt is
based on a density functional theory (DFT),
proposed by Ohta and Kawasaki \cite{OK86}. The DFT leads to the free
energy functional with a nonlocal term measuring the long-range interaction.
Using the time-dependent Ginzburg-Landau theory for equilibrium phase
transition  \cite{MW09a, MW},  a modified Cahn-Hilliard equation
revealing the ODT is obtained, and is also called Ohta-Kawasaki model.

The  mathematical analysis of the model is carried out using  the
dynamic transition theory developed recently  by Ma  and Wang
\cite{MW, MW05}. The main philosophy of this theory is to search for
the full set of  transition states, giving a complete
characterization on stability and  transition. The set of transition
states is often represented by a local attractor. Following this
philosophy, the dynamic transition theory is developed  to identify
the transition states and to classify them both dynamically and
physically.   One important ingredient  of the theory is the
introduction of a new classification scheme of transitions, with
which phase transitions are classified into three types: Type-I,
Type-II and Type-III. In more mathematically intuitive terms,  they
are  called continuous, jump and mixed transitions respectively.
Basically, as the control parameter passes the critical threshold,
the transition states stay in a close neighborhood of the basic
state for a Type-I transition,  are outside of a neighborhood of the
basic state for a Type-II (jump) transition. For the Type-III
transition, a neighborhood is divided into two open regions with a
Type-I transition in one region, and a Type-II transition in the
other region.

We now describe briefly the main results obtained in this article.
Here we indicate only briefly the physical significance of the
results, and more detailed physical conclusions will be addressed in
a separate article.

First, we show that as soon as the linear problem loses stability,
the system always undergoes a dynamic transition to one of the three
types,  forming different patterns/structure. The types of
transition are then dictated by a nondimensional parameter,
measuring the interactions between the long-range repulsive term and
the quadratic and cubic nonlinearities in the model. For example, in
the LAM and HPC cases, this parameter is given by
$$
B\simeq  \gamma_3 -\fraction{8}{27  }\frac{\gamma_2^2}{\sqrt{\sigma}};
$$
see (\ref{B})  and (\ref{B approx}) for the exact formula. $B>0$
indicates first order transitions, and $B<0$ implies second-order
transitions.

Second, the long-range interaction ($\sigma u$ term in the equation)
plays an essential role in pattern selection. In particular, there
are three unique features caused by this long-range interaction
term, which are not present in the phase transition dynamics and
pattern formations described by the classical Cahn-Hilliard model
\cite{MW09a}. {The first one is a novel pattern selection
mechanism associated with the competition between the long-range
repulsive interaction and the short-range attractive interactions,
which leads to patterns like LAM, HPC, etc, as mentioned before.} The
second unique feature is that the scale of the spatial  patterns
emerging from the transition have nanoscale rather than macroscale
periods, determined by the parameter $\sigma$, which measures the
strength of the long-range interaction. Finally, the long-range
interaction causes the periodic structure of the transition
solutions to be very sensitive to the system parameters, which can
be seen from a precise pattern selection criterion (\ref{lambda_c}).

Third, an important technical ingredient of the study is the
reduction of the partial differential equation model to the center
manifold generated by the first unstable modes. Then the reduced
transition equations are a finite-dimensional dynamic system, and
are analyzed carefully following the idea and philosophy of the
dynamic transition theory. Different from many other dynamic
transition problems we have encountered, in the HPC case, the
reduced  transition equations are a  two-dimensional dynamical
system, consisting of (degenerate) quadratic terms and
non-degenerate cubic terms. This unique feature of the reduced
system is caused directly by the introduction of the long-range
interaction term in the model.

Fourth, in the HPC case, the dynamic transition is
a mixed type, Type-III.  The HPC structure is represented by the
local attractor for the continuous transition part of the Type-III
transition. Furthermore, these hexagonal patterns are metastable.
Namely, the original system undergoes a dynamic transition either to
the hexagonal structure or to some more complicated far away
patterns depicted by the jump transition part of the Type-III
transition.

This article is organized as follows. In Section 2, a Cahn-Hilliard
type equation incorporating long range interactions
is derived based on the DFT theory. The linear problem
of this equation is presented in Section 3, and the phase transition
and pattern formation are  studied in Section 4. Section 5 is devoted to the proofs of the main theorems. 

\section{A Dynamic Phase Transition Model for Diblock Copolymer Melts}
Let $u_A$ and $u_B$ be the fraction of monomer number densities of type A and
type B respectively. Since copolymer melt is highly incompressible, we
have $u_A  + u_B = 1$. In the disordered state, $u_A =
a$ is a constant. The free energy functional in terms
of $u_A$ derived from the DFT has the following form:
\begin{align} \label{FreeEnergy}
F(u_A)=& \int_\Omega \fraction{\mu}{2}\,|\nabla \,u_A|^2 + f(u_A) \\
       &+ \fraction{\sigma_d}{2}(-\Delta)^{-\frac{1}{2}}(u_A - a)\cdot(-\Delta)^{-\frac{1}{2}}(u_A -
 a) \, \mathrm{d} x +  F_0.  \nonumber
 \end{align}
Here $F_0$ is a constant, and the coefficients $\mu$ and
$\sigma_d$ are:
 \begin{equation} \label{sigma_d}
 \mu = \frac{{\textstyle kTl^2}}{{\textstyle 12 a(1-a)}}, \; \;  \;  \sigma_d = \frac{{\textstyle 9kT}}{{\textstyle l^2N^2a^2(1-a)^2}},
 \end{equation}
\noindent where $l$ is the Kuhn statistical length as mentioned
before, $k$ is the Boltzman constant, and $N$ is the total number of
monomers in each diblock copolymer chain.

For the energy term $f(u_A)$, we use the following Flory-Huggins formulation for the Gibbs free energy of mixing:
\footnote{In \cite{CR03, NO95}, the
energy term $f(u_A)$ ($W(u_A)$ in their notation)  is a
double well function with global minima at $u_A = 0$ and $u_A=1$,
typically of the form $4u_A ^2(u_A - 1)^2$. The
variable  is $u = u_A - u_B$. Under this notation,
the typical expression of $W(u)$ is $1/4(u^2-1)^2$, which is the one
given in \cite{NO95}.}
\begin{equation} \label{energy of mixing per monomer}
f(u_A) = kT [\fraction{u_A}{N_A}\ln u_A + \fraction{1 - u_A}{N_B}\ln (1 - u_A) + \chi u_A (1-u_A)],
\end{equation}
where $N_A$ and $N_B$ are the number of type-A and type-B monomers
in a diblock copolymer chain respectively, and $\chi$ is the
Flory-Huggins parameter which measures the incompatibility between
type-A and type-B monomers. This expression was first derived to
describe the free energy of polymer solutions and polymer blends
\cite{deG80,FK51,Hug41, Str07}.

The first term in (\ref{FreeEnergy}) represents the interfacial free
energy, the second one measures the bulk energy of the mixing, and
the third term reflects the long range interaction caused by the
connectivity of the monomers in a chain, where $(-\Delta)^{-1/2}$ is
a fractional power of the Laplace operator under zero flux boundary
condition; see \cite{CR03} for more details.

We note that the original formula derived in Ohta and Kawasaki
\cite{OK86} was given for the whole space, the expression on a
bounded domain first appeared in Nishiura and Ohnishi \cite{NO95},
and the exact expression of the coefficients that we adopt here follows
Choksi and Ren \cite{CR03}.

The equation governing the dynamical behavior of the model can be
derived as follows.

Let $J$ be the flux of type-A monomers. Then
\begin{equation*}
J = - m \nabla( \mu _A - \mu _B),
\end{equation*}
where $m$ is the mobility measuring the strength of diffusion (see
\cite{NS84, Cah68}), and $\mu_A$ and $\mu_B$ are the chemical
potentials per monomer of Type-A and Type-B monomers respectively,
which satisfy
\begin{equation*}
 \mu _A - \mu _B = \fraction{\delta}{\delta u_A} F(u_A),
\end{equation*}
where $\delta F(u_A)/\delta u_A$ is the variational derivative of
$F(u_A)$.  See Reichl \cite{Rei98}, Novich-Cohen and Segel
\cite{NS84} and Ma and Wang \cite{MW09a}.

By conservation of mass, we obtain
\begin{equation} \label{Conservation of mass}
\pd{u_A}{t} = - \nabla J = -\nabla (-m\nabla \fraction{\delta
F(u_A)}{\delta u_A}).
\end{equation}

We assume for simplicity that $m$ is a positive constant. Then
(\ref{Conservation of mass}) is reduced to
\begin{equation} \label{Conservation of mass 2}
\pd{u_A}{t} =  m \Delta\fraction{\delta F(u_A)}{\delta u_A}.
\end{equation}

Now let $u = u_A - a$. Using (\ref{FreeEnergy}) and approximating
the $f$ term in \eqref{energy of mixing per monomer} by its Taylor
expansion about $u=0$ up to the fourth order, we obtain the
following equation for $u$ from (\ref{Conservation of mass 2}):
\begin{equation} \label{Eqn. 1}
\begin{aligned}
& \pd{u}{t}  = m [- \mu\Delta^2u + \Delta(b_1u + b_2u^2 +
b_3u^3) -
\sigma_d u ], \\
& \int_\Omega u (x,t) \, \mathrm{d} x  =  0,
\end{aligned}
\end{equation}
where $b_1$, $b_2$, and $b_3$ are as follows:
\begin{equation}
\begin{aligned}
& b_1   =  \fraction{kT}{N}\Bigl [\fraction{a^2 + (1-a)^2}{a^2(1-a)^2} - 2 \chi N \Bigr], \\
& b_2   =  \fraction{kT}{2N} \fraction{a^3 - (1-a)^3}{a^3(1-a)^3}, \\
& b_3   =  \fraction{kT}{3N}\fraction{a^4 + (1-a)^4}{a^4(1-a)^4}.
\end{aligned}
\label{b term}
\end{equation}
The second equation in \eqref{Eqn. 1} is due to the fact that the copolymer melt system has no material exchange with the
external environment. The equation (\ref{Eqn. 1}) is supplemented with Neumann and no-flux
boundary conditions:
\begin{equation} \label{Neumann}
\pd{u}{n} = \pd{\Delta{u}}{n} = 0 \; \; \text{on}\ \partial{{\Omega}}, 
\end{equation}
and for simplicity, we consider the spatial domain to be a bounded
rectangular domain, i.e. $\Omega = \prod_{i=1}^{3}(0, L_i^d)$.

We start with the non-dimensional form of (\ref{Eqn. 1}). Let
\begin{eqnarray*}
\begin{aligned}
x &= dx', & t&= \fraction{d^4}{m \mu}t', && u = u', &&
\lambda = -\fraction{d^2b_1}{\mu}, \\
\gamma_2 &= \fraction{d^2b_2}{\mu}, & \gamma_3&=
\fraction{d^2b_3}{\mu}, && \sigma_d = \fraction{\mu}{d^4}\sigma,
\end{aligned}
\end{eqnarray*}
where $d$ is a typical length scale related to the domain $\Omega$.
By (\ref{sigma_d}) and (\ref{b term}), we have:
\begin{eqnarray}
&& \lambda = \fraction{12d^2a(1-a)}{l^2}\Bigl[2\chi - \fraction{a^2 + (1-a)^2}{Na^2(1-a)^2}\Bigr], \label{lambda} \\
&& \gamma_2 = \fraction{6 d^2 [a^3 - (1-a)^3]}{l^2 N a^2(1-a)^2}, \label{gamma_2} \\
&& \gamma_3 = \fraction{4 d^2 [a^4 + (1-a)^4]}{l^2 N a^3(1-a)^3}, \label{gamma_3} \\
&& \sigma = \fraction{108 d ^4}{l ^4 N ^2 a (1-a)}. \label{sigma}
\end{eqnarray}

With the above non-dimensional variables and parameters, the
equation (\ref{Eqn. 1}) together with the Neumann and no-flux
boundary conditions and initial condition can be rewritten as
follows (omitting the primes):
\begin{equation} \label{Eqn. 2}
\begin{aligned}
& \pd{u}{t}  =  -\Delta ^2u - \lambda\Delta u + \Delta(\gamma_2u^2 + \gamma_3u^3) - \sigma u,\\ 
& \int_\Omega u(x,t) \, \mathrm{d} x = 0, \\ 
& \pd{u}{n} = \pd{\Delta{u}}{n} = 0 \qquad \text{on}\ \partial{{\Omega}},\\ 
& u(0) = \psi, 
\end{aligned}
\end{equation}
where $\Omega = \prod _{i=1}^{3}(0,\, L_i),$ and $L_i =L_i^d/d, 1\leq
i \leq 3.$

For the mathematical set-up, let
\begin{equation*}
\begin{aligned}
H &= \left\{u\in L^2(\Omega) \, \Bigl \vert \: \int_\Omega u \; \mathrm{d} x = 0\, \right\},  \\
H_1 &= \left \{ u\in H^4(\Omega) \cap H \, \Bigl \vert \: \pd{u}{n}
= \pd{\Delta{u}}{n} = 0 \,\, \text{on}\  \partial \Omega \right \}.
\end{aligned}
\end{equation*}

We define the operators $L_\lambda = -A + B_\lambda$ and $G: H_1 \; \to \; H$ by
\begin{equation}
\begin{aligned}
& Au = \Delta^2u, \\
& B_\lambda u = -\lambda \Delta u - \sigma u, \\
& G(u) = \gamma_2\Delta u^2 +\gamma_3 \Delta u^3.
\end{aligned}
\label{Eqn. DefineAbstract}
\end{equation}

Thus, equation (\ref{Eqn. 2}) is equivalent to the following operator equation:
\begin{equation} \label{Eqn. Operator}
\begin{aligned}
&\frac{\mathrm{d} u}{\mathrm{d} t} = L_\lambda u + G(u), \\
& u(0) = \psi. 
\end{aligned}
\end{equation}

It is well known that the Cahn-Hilliard equation can be viewed as a
gradient flow with respect to $H^{-1}$. This is also true for our
equation \eqref{Eqn. 2}. Thus the existence and uniqueness of
solution to \eqref{Eqn. 2} and existence of global attractors can be
proven in a standard fashion.

\section{Principle of Exchange of Stability}
In this section, we consider the eigenvalue problem of the linear part of equation ({\ref{Eqn. Operator}}). Let $\rho_K$, $e_K$ be the
eigenvalues and eigenfunctions of the following eigenvalue problem:
\begin{equation} \label{Eign-VP-Laplacian}
\begin{aligned}
&-\Delta e_K = \rho_K e_K, \\ 
&\pd{e_K}{n} = 0 \quad  \text{ on } \partial \Omega,  \\
&\int_\Omega e_K \mathrm{d} x = 0.
\end{aligned}
\end{equation}

Because of the Neumann boundary condition, $\rho_K$ and $e_K$ must
have the following form:
\begin{equation} \label{e_K}
\begin{aligned}
& \rho_K = \vert K \vert ^2,   \\
& e_K  =  \cos (\fraction{k_1 \pi x_1}{L_1})
\cos ( \fraction{k_2 \pi x_2}{L_2} )\cos
( \fraction{k_3 \pi x_3}{L_3}),
\end{aligned}
\end{equation}
where $K$ is chosen from the following permissible set $\mathcal{P}$:
\begin{equation} \label{K}
\mathcal{P} = \left \{\, (\fraction{k_1 \pi}{L_1} , \,\frac{k_2 \pi}{L_2}, \, \frac{k_3 \pi}{L_3}) \: \Bigl \vert \: k_i \in \overline{\mathbb{Z^-}}, 1\leq i
\leq 3, \, \sum_{i=1}^3 k_i^2 \neq 0 \, \right \},
\end{equation}
and
\begin{equation} \label{|K|}
 \vert K \vert ^2 = \pi^2 \sum_{i=1}^3 k_i^2/ L_i^2.
\end{equation}

It is clear that the linear operator $L_\lambda = -A + B_\lambda$
defined by (\ref{Eqn. DefineAbstract}) with (\ref{Neumann}) has the same eigenfunctions $\{e_K \:|\: K \in \mathcal{P} \}$ as in (\ref{e_K}). The eigenvalue of $L_\lambda$ corresponding
to each $e_K$ is given by
\begin{equation} \label{Eigenvalues}
 \beta_K(\lambda) = - \vert K \vert ^4 + \lambda \vert K \vert ^2 - \sigma = \vert K \vert ^2(\lambda - \fraction{\vert K \vert ^4 + \sigma}{\vert K \vert ^2}).
\end{equation}

The linear stability and instability are precisely determined by the
critical-crossing of the first eigenvalue, which is often called
principle of exchange of stabilities (PES). For this purpose, we
define a number $\lambda_c$ by the following:
\begin{equation} \label{lambda_c}
\lambda_c = \min_{K \in \mathcal{P}} \fraction{\vert K \vert ^4 + \sigma}{\vert K \vert ^2}.
\end{equation}
Now, let
\begin{equation} \label{set S}
\mathcal{S} = \left \{\, K \in \mathcal{P} \, \vert \, K \text{ achieves the minimum in \eqref{lambda_c}} \,\right \},
\end{equation}
then the following PES condition holds true:
\begin{align} \label{PES}
&
 \beta_{K}(\lambda)  \left\{  \begin{aligned}
   & <0  &&  \text{ if } \lambda <\lambda_c,\\
   & =0  &&  \text{ if } \lambda =\lambda_c,\\
   & >0  &&   \text{ if } \lambda >\lambda_c
\end{aligned}
\right.
    && \text{for } K \in \mathcal{S}, \\
&
\beta_{K}(\lambda_c)<0
     &&  \text{for all } K \in \mathcal{P} \backslash \mathcal{S}.
\end{align}

The above PES condition shows that $\lambda_c$ is exactly the
critical value of $\lambda$, such that the disordered state $u=0$ is
linearly stable when $\lambda < \lambda_c$, and becomes linearly
unstable when $\lambda$ crosses $\lambda_c$ from below.

Note that $\lambda_c$ is achieved at some $K_c\in \mathcal{S}$ such
that $|K_c|^2 \approx \sqrt{\sigma}$. By \eqref{sigma}, this can be
written in the dimensional form as:
\begin{equation} \label{critical K eqn}
|K_c| = \left( \sum_{j=1}^3 \frac{(k_{j}^c\pi)^2}{(L_j^d)^2} \right)^{1/2} \approx
\frac{\sqrt[4]{108}}{l \sqrt{N}\sqrt[4]{a(1-a)}},
\end{equation}
where $L_j^d (1\le j \le 3)$ is the dimensional size of the domain. Since the Kuhn statistical length $l$ is of molecular length scale,
from \eqref{critical K eqn} we can see that structure of high
spatial frequency appears naturally when the disordered state
becomes unstable, and the critical wave vector
$(\frac{k_1^c\pi}{L_1}, \frac{k_2^c\pi}{L_2}, \frac{k_3^c\pi}{L_3})$
depends sensitively on system parameters.

\section{Dynamic Phase Transition and Pattern Formation} \label{main thms}
By the dynamic transition theory \cite{MW, MW05}, the we know that
as $\lambda$ crosses $\lambda_c$ from below, the system always undergoes a dynamic transition to one of the three types, and the type of transition is dictated by the nonlinear interactions.
In this section, we address the nonlinear interaction, the type and structure of the phase transitions, and the associated pattern formations.

\subsection{Transitions to LAM Patterns} In this section, we consider the case where  the first
eigenvalue is simple and the corresponding $K_1$ which satisfies
(\ref{lambda_c}) is of the form $K_1=(k_1\pi/L_1,\ 0,\ 0)$,
$k_1 \neq 0$.  The type and structure of phase transitions of the system is dictated by the sign of the following parameter:
\begin{equation} \label{B}
B= \gamma_3 - \fraction{8 \vert K_1 \vert ^2}{36 \vert K_1 \vert ^4 - 9\sigma}\gamma_2^2.
\end{equation}
As mentioned before,
\begin{equation}\label{approx}
|K_1|^2 \simeq \sqrt{\sigma}, \qquad \lambda_c\simeq 2\sqrt{\sigma}.
\end{equation}
Hence the parameter $B$  takes the following form:
\begin{equation} \label{B approx}
B\simeq  \gamma_3 -\fraction{8}{27  }\frac{\gamma_2^2}{\sqrt{\sigma}}.
\end{equation}

\begin{figure}
  \centering
   \includegraphics[height=3.5cm, width=3.5cm]{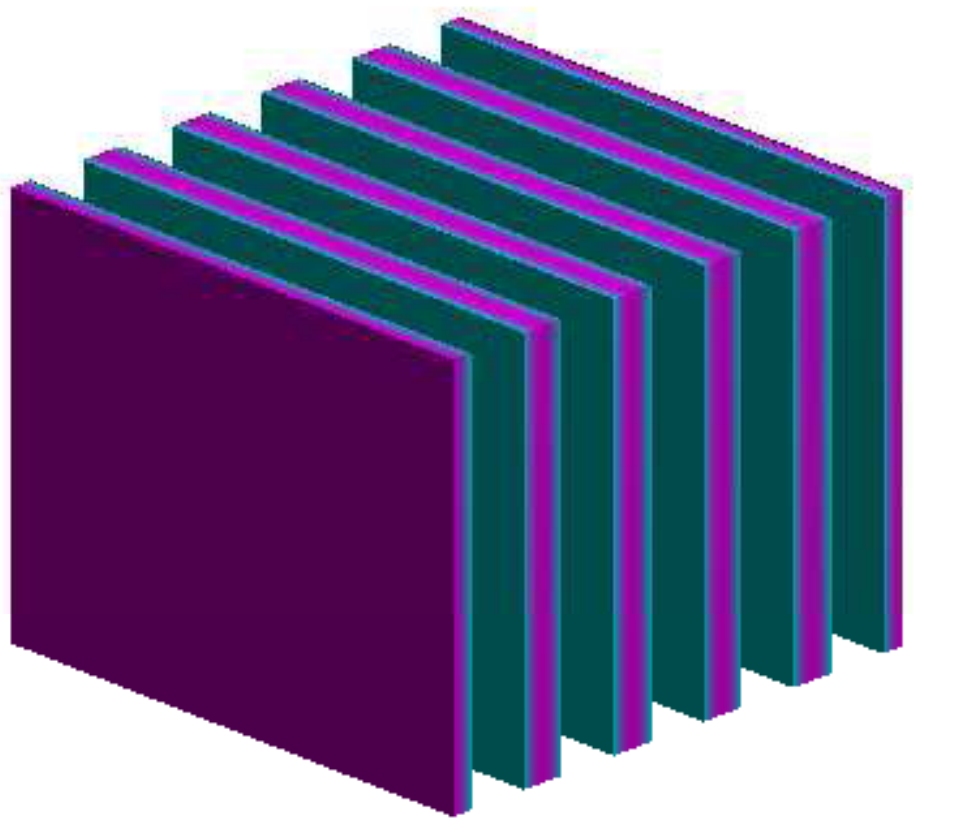}
\caption{LAM pattern: a schematic structure of $u_{1,2}$ given in \eqref{attractor_laminar}.}
\label{laminar}
\end{figure}

\bt \label{bif. thm} When $K_1=(k_1\pi/L_1,\ 0,\ 0)$, $k_1
\neq 0$, is the only wave vector which satisfies (\ref{lambda_c}),
the following assertions hold true:

\begin{itemize}
\item[(1)] If
$
B < 0,
$
then the phase transition of (\ref{Eqn. 2}) at
$\lambda_c$ is Type-II, where $\lambda_c = (|K_1|^4 + \sigma ) / |K_1|^2$ and $|K_1|^2 = k_1^2\pi^2/L_1^2$. In particular, the problem bifurcates on
the side $\lambda < \lambda_c$ to two non-degenerate saddle points,
and there are two saddle-node bifurcations at some $\lambda_\ast < \lambda_c$ as shown in Figure
\ref{saddle-node}.

\begin{figure}
   \subfigure[ ]{
\label{saddle-node}
   \begin{minipage}[b]{0.4\textwidth}
   \centering
   \includegraphics[height=4.5cm, width=4.5cm]{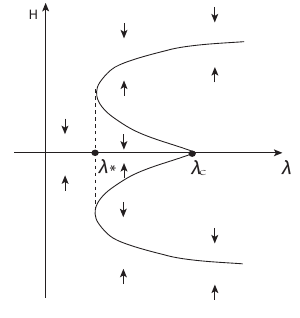}
\end{minipage}}
   \subfigure[ ]{
   \begin{minipage}[b]{0.4\textwidth}
   \centering
   \includegraphics[height=4.5cm, width=4.5cm]{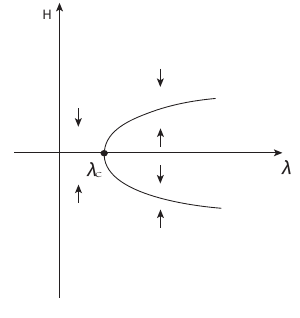}
\label{type-I}
\end{minipage}}
  \caption{(a) Type-II transition given by Theorem \ref{bif. thm}: The disordered state $u=0$ is globally stable when $\lambda < \lambda_{\ast}$, is metastable when $\lambda_{\ast}< \lambda <  \lambda_c$, and becomes unstable when $\lambda > \lambda_c$. There are two saddle-node bifurcations at $\lambda_\ast$. (b) Tyep-I transition as given by Theorem \ref{bif. thm}.}
\end{figure}

\item[(2)]  If $B > 0$,
then the transition is Type-I, and the problem bifurcates
on $\lambda > \lambda_c$ to two attractors $u_1$ and $u_2$ as shown in Figure \ref{type-I}, which are given by (see Figure \ref{laminar} for a structure of $u_{1,2}$):
\begin{equation} \label{attractor_laminar}
u_{1,2} = \pm \sqrt{\fraction{4\beta_{K_1}}{3\sqrt{\sigma} B}} \cos(\fraction{k_1\pi x_1}{L_1}) + o(\vert \beta_{K_1} \vert ^{1/2}),
\end{equation}
\noindent where $\beta_{K_1}$ is given in (\ref{Eigenvalues}).
\end{itemize}
\et

Two remarks are now in order.

 \medskip

 {\sc First},  as we know, a Type-II transition correspond to the first-order transition in the sense of Ehrenfest. In this case,   $\lambda_\ast$ and $\lambda_c$ in Figure \ref{saddle-node}
represent respectively the so called binodal and spinodal points. From (\ref{lambda_c}), we know that $\lambda_c$ is approximately $2\sqrt{\sigma}$, then by (\ref{lambda}) and (\ref{sigma}), we have the following formula for the spinodal which involves only $\chi N$ and
$a$ as independent parameters:
\begin{equation*}
\chi N \approx \fraction{\sqrt{3}}{2(a(1-a))^{3/2}} + \fraction{a^2 + (1-a)^2}{2a^2(1-a)^2}.
\end{equation*}
To our knowledge, there is
no such simple formula for the binodal point. Nevertheless, some
efforts have been devoted to figuring out $\lambda_\ast$ numerically
e.g. in \cite{CPW09}.

In addition, when the control parameter $\lambda$ is between
$\lambda_\ast$ and $\lambda_c$,  the disordered state $u=0$ is 
metastable. Perturbation may lead the system to other metastable
states which may be far away from the original disordered one.

\medskip

{\sc Second}, Type-I transition corresponds to the second-order or
higher-order transition in physics. In this case the transition
states are determined by the perturbation of the first eigenvectors.
In Theorem \ref{bif. thm}, the transition states are  $u_{1,2}$,
which are stable states, and represent a lamellar pattern as shown
in Figure~\ref{laminar}.

\subsection{Transitions to HPC Patterns.}
Now, we consider the case when the domain size satisfies
the following condition:
\begin{equation} \label{hex condition}
L_1 = 2\pi L, \: L_2 = \frac{2}{\sqrt{3}}\pi L, \text{ and } L_3 =\theta \pi L.
\end{equation}
Here $L$ and $\theta$ are some positive constants, which are chosen in a way such that $K_1^c:=(\frac{n}{L}, 0, 0)$ and $K_2^c := (\frac{n}{2L},\frac{\sqrt{3}n}{2L},0)$ are the only wave vectors satisfying \eqref{lambda_c}, where $n$ is a positive integer.

\begin{figure}
    \subfigure[]{
    \label{attractor hex case}
    \begin{minipage}[b]{0.3\textwidth}
    \centering
   \includegraphics [height=0.85\textwidth, width=1\textwidth, trim = 0 0 0 0, clip]{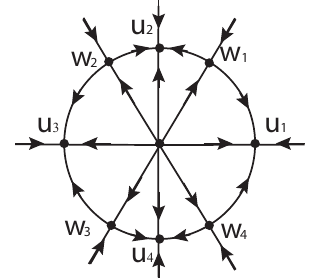}
    \end{minipage}}
    \subfigure[]{
    \label{type III attractor a>0}
    \begin{minipage}[b]{0.3\textwidth}
    \centering
   \includegraphics [height=1\textwidth, width=1\textwidth, trim = 0 0 0 0, clip]{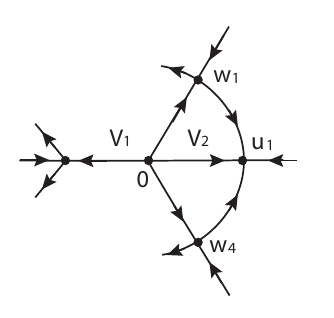}
    \end{minipage}}
    \subfigure[]{
    \label{type III attractor a<0}
    \begin{minipage}[b]{0.3\textwidth}
    \centering
   \includegraphics [height=1\textwidth, width=1\textwidth, trim = 0 0 0 0, clip]{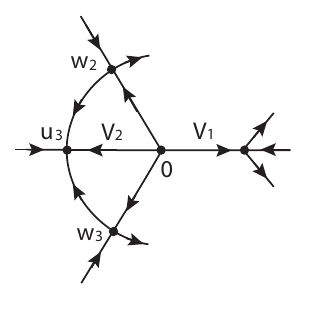}
    \end{minipage}}
  \caption{(a) For $\gamma_2=0$, the attractor $\Sigma_\lambda$ after the transition is homeomorphic to
  $S^1$ and $u_1, u_2, u_3$ and $u_4$ are minimal attractors.
  (b) For $\gamma_2 > 0$, the sectorial regions $V_1$ and $V_2$ are separated by the lines $\overline{0w_1}$ and
  $\overline{0w_4}$, and $u_1$ is the minimal attractor in $V_2$.
  (c) For $\gamma_2 < 0$, the sectorial regions $V_1$ and $V_2$ are separated by the lines $\overline{0w_2}$ and
  $\overline{0w_3}$, and $u_3$ is the minimal attractor in region
  $V_2$.}
\end{figure}

\bt \label{bif. thm hex} Assume that the size of the domain
satisfies (\ref{hex condition}) and $K_1^c=(\frac{n}{L}, 0, 0)$ and
$K_2^c = (\frac{n}{2L},\frac{\sqrt{3}n}{2L},0)$ are the only two
wave vectors which satisfy (\ref{lambda_c}). Let $B$ be the parameter defined in (\ref{B}) with $|K_1| = |K_1^c|=|K_2^c|$. The following assertions hold true:
\begin{itemize}
\item[(1)] If
$$\gamma_2=0,$$
then the phase transition of (\ref{Eqn. 2}) at
$\lambda_c$ is Type-I, where $\lambda_c$ is given by
(\ref{lambda_c}). The problem bifurcates on the side $\lambda
> \lambda_c$ to an attractor $\Sigma_\lambda$, which is homeomorphic
to the one-dimensional unit sphere $S^1$. $\Sigma_\lambda$ contains
eight non-degenerate steady states, with four saddle points $w_1$,
$w_2$, $w_3$ and $w_4$ and four minimal attractors $u_1$,$u_2$,
$u_3$ and $u_4$ as shown in Figure \ref{attractor hex case}.
Moreover, we have the following approximation formulas for the
steady states:
\begin{equation} \label{attractor formulas}
\begin{aligned}
u_{1,3} =& \pm \sqrt{-\frac{\beta_1(\lambda)}{b}}\cos(\frac{n}{d}x_1) + o(|\beta_1|^{1/2}), \\
u_{2,4} =& \pm
\sqrt{-\frac{\beta_1(\lambda)}{d}}\cos(\frac{n}{2d}x_1)\cos(\frac{\sqrt{3}n}{2d}x_2)
+ o(|\beta_1|^{1/2}), \\
w_{1,2,3,4} =& \pm
\sqrt{-\frac{\beta_1(\lambda)}{b+4c}}\cos(\frac{n}{d}x_1) \\
&\pm 2
\sqrt{-\frac{\beta_1(\lambda)}{b+4c}}\cos(\frac{n}{2d}x_1)\cos(\frac{\sqrt{3}n}{2d}x_2)
+ o(|\beta_1|^{1/2}),
\end{aligned}
\end{equation}
where $b$ and $d$ are given in (\ref{b-e}) and $\beta_1(\lambda) =
\beta_{K^c_1}$ is as in (\ref{Eigenvalues}).

\item[(2)]  If
\begin{displaymath}
\gamma_2 \neq 0 \text{ and } B<0,
\end{displaymath}
then the problem bifurcates on both sides of $\lambda_c$ and the
transition is Type-II. Moreover, there are four steady states
bifurcated out on the side $\lambda < \lambda_c$, including three
saddle points and one unstable node. On the side $\lambda
> \lambda_c$, the problem bifurcates to two steady states, which are saddles.

\item[(3)]  If
\begin{displaymath}
\gamma_2 \neq 0 \text{ and } B>0,
\end{displaymath}
then the transition is Type-III. Again, there are bifurcations on
both sides of $\lambda_c=2\sqrt{\sigma}$. On the side $\lambda <
\lambda_c$, there are two saddles bifurcating out from the origin.
On the side $\lambda
> \lambda_c$, the problem bifurcates to four steady states, including one stable node and three saddles. Moreover, there is a neighborhood $V\subset H$ of $u=0$, which can be decomposed into two
disjoint sectorial regions $V_1$ and $V_2$ such that $\overline{V}=
\overline{V}_1 \cup \overline{V}_2$ and the phase transition is
first order in $V_1$ and is $n$-th order in $V_2$ with $n \ge 2$ in
the Ehrenfest sense. $V_1$ and $V_2$ satisfy the relation given in (\ref{V}). In region $V_2$, there is exactly one minimal
attractor as is shown in Figure \ref{type III attractor a>0} for
$\gamma_2 > 0$ and in Figure \ref{type III attractor a<0} for
$\gamma_2 <0$. The minimal attractors $u_1$ and $u_3$ can again be
approximated as in (\ref{attractor formulas}).
\end{itemize}
\et

\begin{figure}
    \subfigure[]{
    \begin{minipage}[b]{0.4\textwidth}
    \centering
   \includegraphics [height=0.85\textwidth, width=1\textwidth, trim = 0 0 0 0, clip]{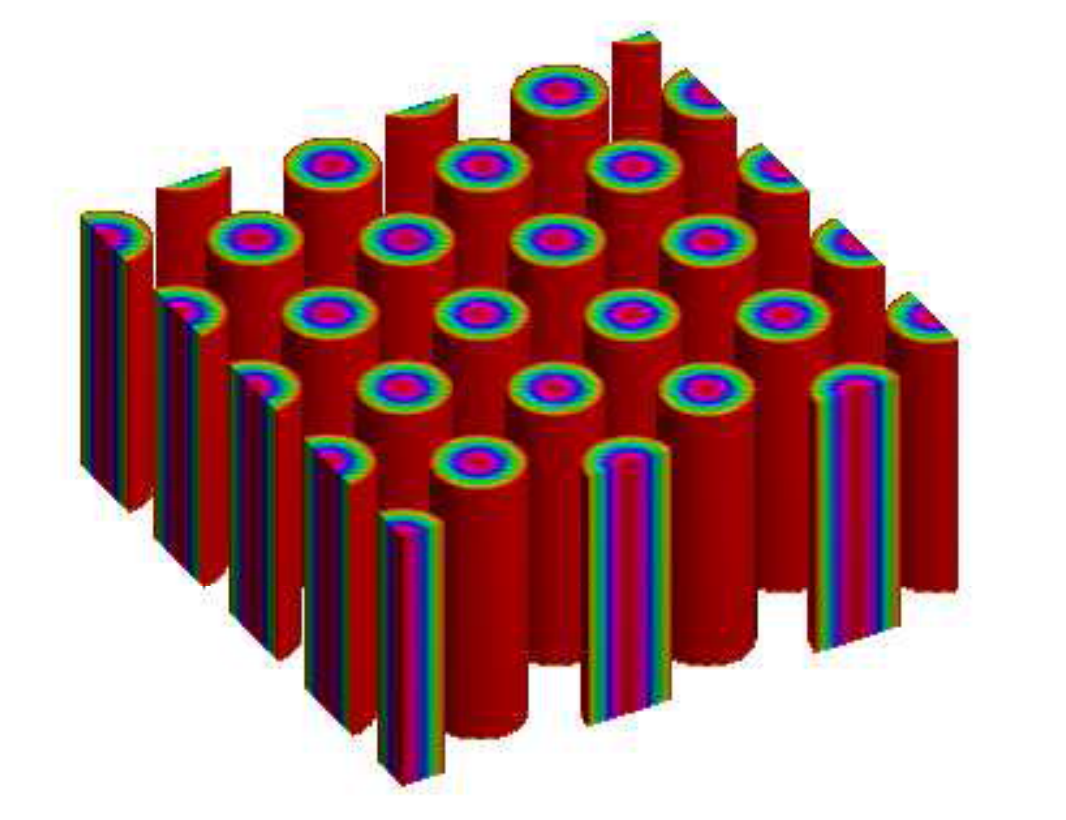}
    \end{minipage}}
    \subfigure[]{
    \begin{minipage}[b]{0.4\textwidth}
    \centering
   \includegraphics [height=1\textwidth, width=1\textwidth, trim = 0 0 0 0, clip]{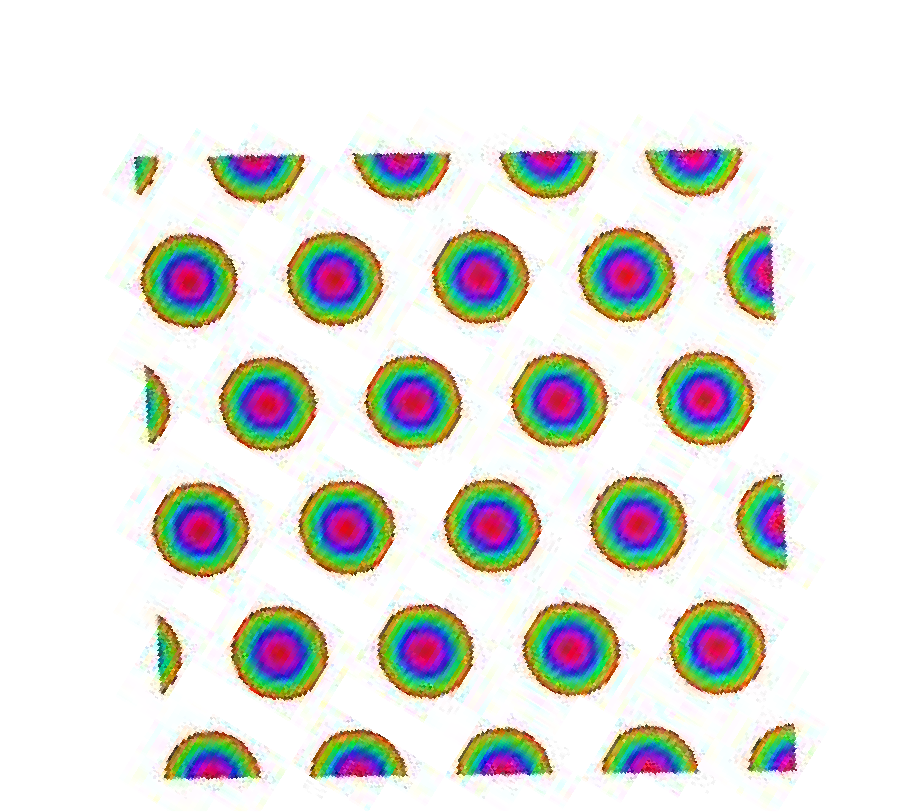}
    \end{minipage}}
\caption{(a) HPC structures determined by $w_{1,2,3,4}$ in \eqref{attractor formulas}.  (b) Top view of the structures given in part (a).}
\label{hexagon}
\end{figure}

We note that, among all the steady states given in
\eqref{attractor formulas}, $u_{1,3}$ are related to LAM as shown in
Figure \ref{laminar}, $w_{1,2,3,4}$ lead to HPC given in Figure
\ref{hexagon}, and $u_{2,4}$ correspond to rectangular patterns as
in Figure \ref{rectangle}. In the case of Type-I transition, as the
disordered state loses its stability, the transition happens in two
stages. There is first a fast transition from the disordered state to
the bifurcated attractor $\Sigma_\lambda$, then followed by a slow
transition within the bifurcated structure. Depending on the initial
perturbations, the patterns finally emerging from the transition may
be either lamellae or rectangles.

In Type-III transition case, transition may happen on both
sides of the critical point $\lambda_c$. On the side $\lambda <
\lambda_c$, the disordered state is metastable, and perturbations
may lead the system to some other metastable states far away from
the disordered state as discussion in last section. On the side
$\lambda > \lambda_c$, the disordered state is unstable, and the
transition may be either first order or second order depending on
the initial perturbations. When the perturbation leads to second
order transition, there is first a fast transition to the local
attractor composed with two steady states with the HPC structure,
one steady state with LAM structure, and two heteroclinic orbits
connecting them as shown in Figures \ref{type III attractor a>0} and
\ref{type III attractor a<0}. The pattern eventually goes to LAM.

\subsection{Transitions to Squares and Spheres}
In this section, we return to the situation when the first eigenvalue is
simple, and we study the case when the corresponding $K_1$
satisfying (\ref{lambda_c}) is of the form $K_1=(k_1\pi/L_1, \
k_2\pi/L_2,\ 0)$, $k_i\neq 0$, $1 \leq i \leq 2$, or
$K_1=(k_1\pi/L_1, \ k_2\pi/L_2,\ k_3\pi/L_3)$, $k_i\neq
0$, $1 \leq i \leq 3$. For simplicity, we will give results for the case $k_1/L_1=k_2/L_2=k_3/L_3$, but the general situation can be dealt with in the same way. With this assumption, the first critical eigenvectors will have square pattern when $K_1=(k_1\pi/L_1, \
k_2\pi/L_2,\ 0)$, and a cubic pattern when
$K_1=(k_1\pi/L_1, \ k_2\pi/L_2,\ k_3\pi/L_3)$.
We define the following parameters:
\begin{equation} \label{B2,B3}
\begin{aligned}
& B_2= \gamma_3 -\frac{16}{9}\left(\frac{\vert K_1 \vert^2}{2\vert K_1 \vert^4-\sigma}+\frac{1}{2} \fraction{\vert K_1 \vert^2}{14 \vert K_1 \vert^4 - 2 \lambda \vert K_1 \vert^2 - \sigma}\right)\gamma_2^2,\\
& B_3=\gamma_3-\frac{32\vert K_1\vert^2}{3}\biggl(
\frac{1}{-2\vert K_1\vert^4+6\lambda\vert K_1\vert^2-9\sigma}+\frac{1}{46\vert K_1\vert^4-6\lambda\vert K_1\vert^2-9\sigma}\\
&\hspace{1.5in} +\frac{1}{252\vert K_1\vert^4-108\lambda\vert K_1\vert^2-18\sigma} 
\biggr)\gamma_2^2.
\end{aligned}
\end{equation}
Using \eqref{approx}, we have
\begin{equation} \label{B2 B3 approx}
\begin{aligned}
& B_2\simeq  \gamma_3 -\fraction{152}{81  }\frac{\gamma_2^2}{\sqrt{\sigma}},\\
& B_3\simeq  \gamma_3 -(11+\fraction{463}{675})\frac{\gamma_2^2}{ \sqrt{\sigma}}. 
\end{aligned}
\end{equation}

\begin{figure}
   \begin{minipage}[t]{0.48\textwidth}
   \centering
   \includegraphics[height=3.5cm, width=3.5cm]{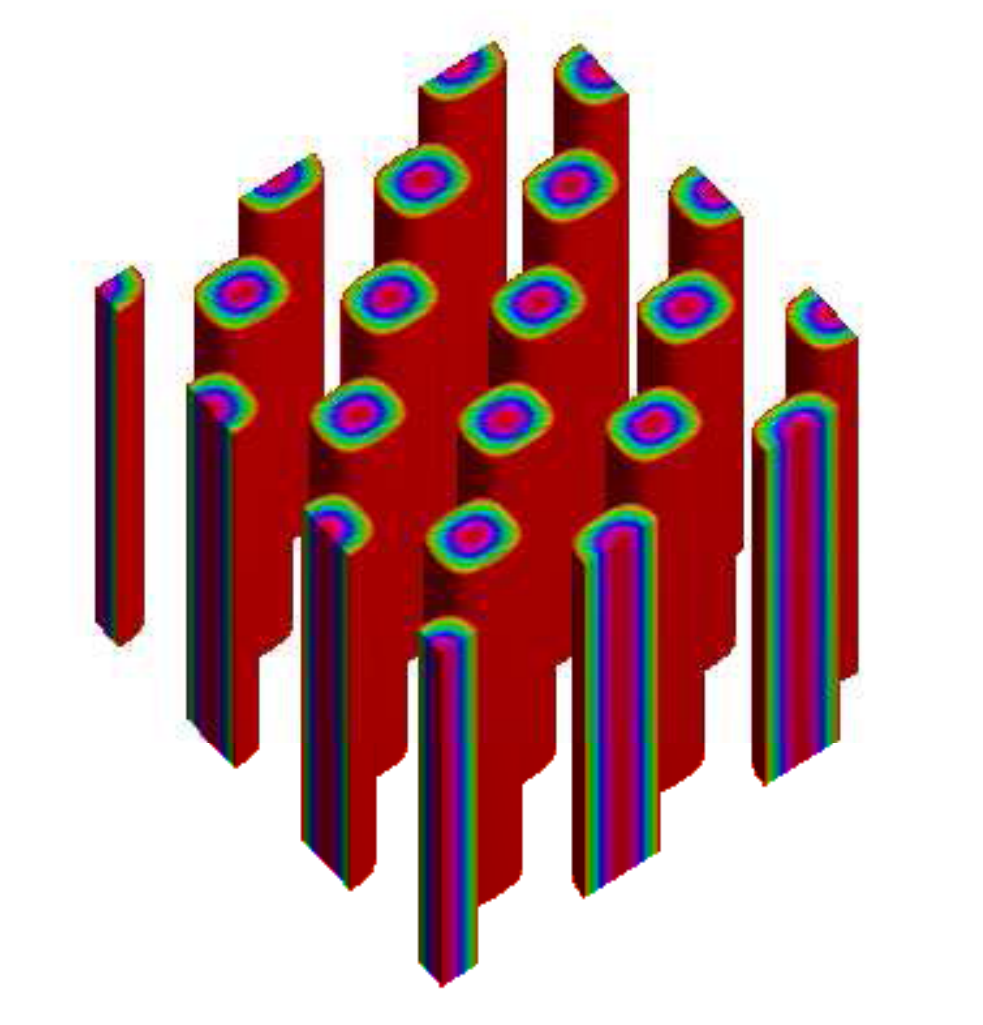}
\setcaptionwidth{0.8\textwidth}
\caption{Rectangular patterns: a schematic structure of $u_{1,2}$ given in \eqref{attractor_rectangle}.}
\label{rectangle}
\end{minipage}
   \begin{minipage}[t]{0.48\textwidth}
   \centering
   \includegraphics[height=3.5cm, width=3.5cm]{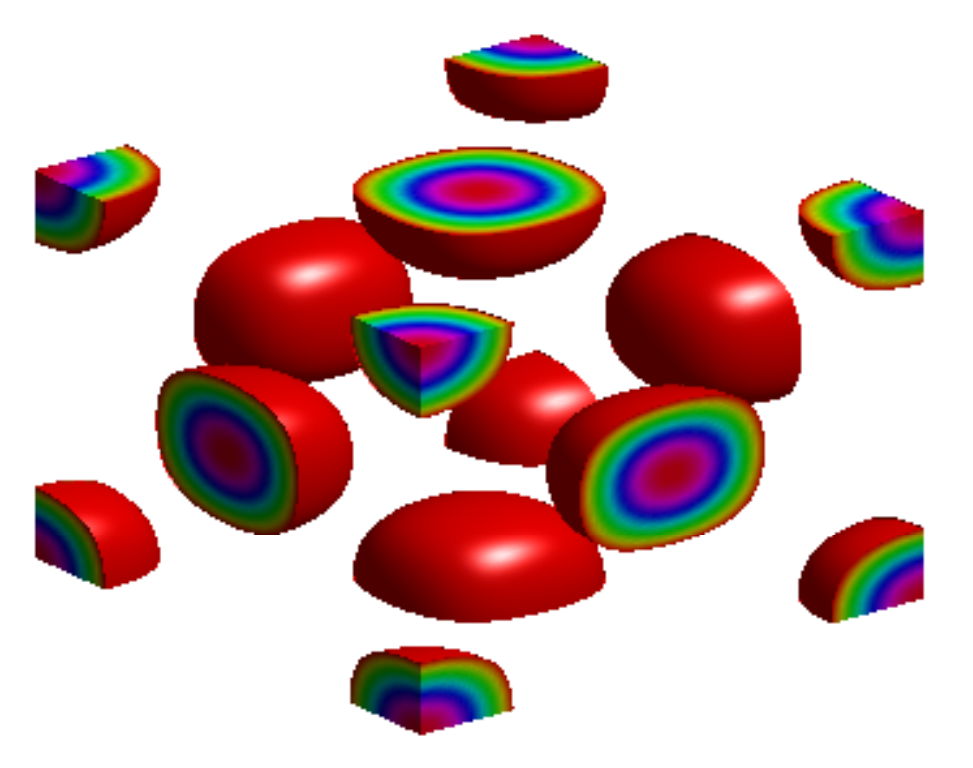}
\setcaptionwidth{0.8\textwidth}
\caption{Sphere patterns in face-centered-cubic lattices: a schematic structure of $u_{1,2}$ given in \eqref{attractor_cube}.}
  \label{cube}
\end{minipage}
\end{figure}

\bt \label{bif. thm case 1.2}
Assume that $K_1=(k_1\pi/L_1, \ k_2\pi/L_2,\ 0)$, $k_1/L_1=k_2/L_2\neq0$, is the only wave vector which satisfies (\ref{lambda_c}). Then the following assertions hold true:

\begin{itemize}
\item[(1)] If
\begin{displaymath}
B_2<0,
\end{displaymath}
then the phase transition of (\ref{Eqn. 2}) at
$\lambda_c$ is Type-II, where the critical value $\lambda_c$ is
given by (\ref{lambda_c}). In particular, the problem bifurcates
from $(u,\, \lambda)=(0,\, \lambda_c)$ on the side $\lambda<
\lambda_c$ to two non-degenerate saddle points, and there are two
saddle-node bifurcations at some $\lambda_\ast < \lambda_c$.

\item[(2)]  If
\begin{displaymath}
B_2>0,
\end{displaymath}
the transition is Type-I, and the problem bifurcates
on $\lambda > \lambda_c$ to two attractors $u_1$ and $u_2$, which can be expressed as (see Figure \ref{rectangle} for the structure of $u_{1,2}$)
\begin{align} \label{attractor_rectangle}
u_{1,2} = \pm \sqrt{\fraction{16\beta_{K_1}}{9\sqrt{\sigma}B_2}} \cos(\fraction{k_1\pi x_1}{L_1})\cos(\fraction{k_2\pi x_2}{L_2}) + o(\vert \beta_{K_1} \vert ^{1/2}),
\end{align}
where $\beta_{K_1}$ is as in (\ref{Eigenvalues}).
\end{itemize}
\et

\bt \label{bif. thm case 1.3}
Assume that the only wave vector which satisfies (\ref{lambda_c}) is $K_1=(k_1\pi/L_1, \ k_2\pi/L_2,\ k_3\pi/L_3)$, $k_1/L_1=k_2/L_2=k_3/L_3\neq0$. 
Then the following assertions hold true:

\begin{itemize}
\item[(1)] If
\begin{displaymath}
B_3<0,
\end{displaymath}
then the phase transition of (\ref{Eqn. 2}) at
$\lambda_c$ is Type-II, where the critical value $\lambda_c$ is
given by (\ref{lambda_c}). In particular, the problem bifurcates on
the side $\lambda< \lambda_c$ to two non-degenerate saddle points,
and there are two saddle-node bifurcations.

\item[(2)]  If
\begin{displaymath}
B_3 > 0,
\end{displaymath}
the transition is Type-I, and the problem bifurcates
on $\lambda > \lambda_c$ to two attractors $u_1$ and $u_2$, which can be expressed as (see Figure \ref{cube} for the structure of $u_{1,2}$)
\begin{equation} \label{attractor_cube}
u_{1,2} = \pm \sqrt{\fraction{64\beta_{K_1}}{27\sqrt{\sigma}B_3}} \cos(\fraction{k_1\pi x_1}{L_1})\cos(\fraction{k_2\pi x_2}{L_2})\cos(\fraction{k_3\pi x_3}{L_3}) + o(\vert \beta_{K_1} \vert ^{1/2}),
\end{equation}
\noindent where $\beta_{K_1}$ is as in (\ref{Eigenvalues}).
\end{itemize}
\et

\section{Proofs of Main Theorems}
\bp[Proof of Theorem \ref{bif. thm}] The proof relies on the
following center manifold reduction, which reduce equation
(\ref{Eqn. 2}) to a one-dimensional ODE.

Let $u = v + \Phi(y,\lambda)$, where $v = y e_{K_1}$, and $\Phi$ is
the center manifold function which we will calculate later. Thanks
to the classical center manifold theorem \cite{Hen81, Tem97, MW05},
we have $\langle v, \, \Phi \rangle _H = 0$ and $\Phi(y, \lambda)=o(
\vert y \vert)$. Now multiplying both sides of (\ref{Eqn. 2}) by
$e_{K_1}$ and integrating over $\Omega$, we obtain the reduced
equation to the center manifold:
\begin{equation} \label{Bif. Eqn.}
\fraction{\mathrm{d} y}{\mathrm{d} t}  =  \beta_{K_1} (\lambda) y -
\fraction{2k_1^2 \pi^2}{L_1^2 \vert \Omega \vert}g(y) + o(\vert y
\vert ^3),
\end{equation}
\noindent where $\vert \Omega \vert$ is the volume of $\Omega$, and
{\setlength\arraycolsep{2pt}
\begin{eqnarray}
g(y) & = & G_2(y) + G_3(y) + G_{23}(y), \label{g term}  \\
G_2(y) & = & \gamma_2 \int_{\Omega} v^2 e_{K_1} \, \mathrm{d} x, \label{G2 term} \\
G_3(y) & = & \gamma_3 \int_{\Omega} v^3 e_{K_1} \, \mathrm{d} x, \label{G3 term}\\
G_{23}(y) & = & 2 \gamma_2 \int_{\Omega} v\Phi(y,\,\lambda) e_{K_1}
\, \mathrm{d} x. \label{G23 term}
\end{eqnarray}}
By direct computation, we know
\begin{eqnarray}
&G_2  =  \int_{\Omega}(y\cos (k_1 \pi x_1)/L_1)^2 \cos (k_1\pi x_1/L_1)\,\mathrm{d} x = 0,  \label{G2 term compute} \\
&G_3  = \int_{\Omega} (y \cos(k_1\pi x_1/L_1))^3 \cos (k_1\pi
x_1/L_1)\,\mathrm{d} x = \fraction{3\gamma_3}{8}\vert \Omega \vert
y^3. \label{G3 term compute}
\end{eqnarray}

To evaluate $G_{23}$, we need to compute the center manifold
function $\Phi(y,\,\lambda)$.  We will use the following second order
approximation formula of $\Phi(y,\, \lambda)$ (see e.g. \cite{MW} for details):
\begin{equation} \label{abstract cmd app}
-\mathcal{L}_\lambda \Phi(y,\lambda) = P_2G_2(y, \lambda) + O(|\beta(\lambda)||y|^2) + o(|y|^2).
\end{equation}
Here $P_2: H \rightarrow H_2^\lambda$ is the canonical projection with $H_2^\lambda$ being the subspace of $H$ spanned by all stable eigenfunctions of $L_\lambda$, $\mathcal{L}_\lambda$ is the restriction of $L_\lambda$ to $H_2^\lambda$, and $\beta(\lambda)$ is the first eigenvalue of $L_\lambda$.

The above formula \eqref{abstract cmd app} implies that
\begin{align*}
\Phi(y,\lambda) &= \sum_{ K\in \mathcal{P}, K \neq K_1} \Phi _K
(y,\lambda )e_K  + o(\vert y \vert ^2), \quad \text{where} \\
\Phi _K (y,\lambda) & =  \fraction{\gamma_2}{(2\beta_{K_1} (\lambda)
- \beta_K )\langle e_K, \,
e_K \rangle _H}\int_ {\Omega} \Delta v^2 e_K \, \mathrm{d} x \\
 &=  - \fraction{\gamma _2 \rho_K}{(2 \beta_{K_1} (\lambda) - \beta_K) \langle e_K, \, e_K \rangle _H} \int_ {\Omega} v^2 e_K \, \mathrm{d} x.
\end{align*}

Note that
\begin{displaymath}
\int_{\Omega} e_{K_1}^2e_K \, \mathrm{d} x =  \int_{\Omega} \fraction{1+\cos(2k_1\pi x_1/L_1 )}{2}e_K \, \mathrm{d} x \\
= 0 \text{\: unless \:} K = (2k_1\pi/L_1, \, 0, \, 0).
\end{displaymath}

For $K= (2k_1\pi/L_1, \, 0, \, 0)$, we have:
\begin{equation} \label{Phi_k}
\Phi_K(y,\lambda) = -2 \xi \gamma_2y^2,
\end{equation}
\noindent where $\xi =
\fraction{k_1^2\pi^2/L_1^2}{14k_1^4\pi^4/L_1^4-2\lambda
k_1^2\pi^2/L_1^2-\sigma}$.

Thus we obtain the center manifold function as follows:
\begin{align} \label{CMF}
\Phi(y,\lambda) = -2 \xi \gamma_2 y^2 \cos(\fraction{2k_1\pi
x_1}{L_1}) + o(\vert y \vert ^2).
\end{align}

Plugging (\ref{CMF}) into (\ref{G23 term}), we find
\begin{eqnarray} \label{G23 term compute}
G_{23}= - \xi \vert \Omega \vert \gamma ^2_2 y^3 + o(\vert y \vert
^3).
\end{eqnarray}

Finally, plugging (\ref{g term}), (\ref{G2 term compute}), (\ref{G3
term compute}) and (\ref{G23 term compute}) into (\ref{Bif. Eqn.}),
we can derive the following reduced equation of (\ref{Eqn. 2}):
\begin{align} \label{Bif. Eqn. simplified}
\frac{\mathrm{d} y}{\mathrm{d} t}   =  \beta_{K_1} (\lambda) y -
\fraction{2k_1^2 \pi ^2}{L_1^2}\Bigl(\fraction{3}{8}\gamma_3 -
\frac{k_1^2\pi^2/L_1^2}{\frac{14k_1^4\pi^4}{L_1^4}-\frac{2\lambda
k_1^2\pi^2}{L_1^2}-\sigma} \gamma_2^2 \Bigr)y^3 + o(\vert y \vert ^3).
\end{align}

It is known from \cite{MW, MW05} that the transition type of
(\ref{Eqn. 2}) at the critical point $\lambda_c$ is completely
determined by (\ref{Bif. Eqn. simplified}). Thus all the assertions
of the theorem follow from (\ref{Bif. Eqn. simplified}) except the
claim for the saddle-node bifurcation in Assertion (1), which we
prove as follows.

When $\gamma_3 < \fraction{8 \vert K_1 \vert ^2}{36 \vert K_1
\vert ^4 - 9\sigma}\gamma_2^2$, we
know from (\ref{Bif. Eqn. simplified}) that there are two branches
of saddle points bifurcating from $(u,\,\lambda)=(0,\, \lambda_c)$.
The result will follow if we can show that there exists
$\widetilde{\lambda} < \lambda_c$ such that the following two
conditions hold:
\begin{enumerate}
\item[(i)] $u=0$ is globally stable for any $\lambda < \widetilde{\lambda}$.
\item[(ii)] The two branches of saddle points are uniformly bounded when $\widetilde{\lambda} < \lambda < \lambda_c$.
\end{enumerate}

To verify (i), we do an energy estimate. For any $\lambda <
\widetilde{\lambda}=-\gamma_2^2/2\gamma_3$, and $u \neq 0$ in $H$,
multiplying the right hand side of (\ref{Eqn. 2}) by $u$, and integrating
over $\Omega$, we get:
\begin{equation*}
\begin{aligned}
&\int_\Omega -\vert \Delta u \vert ^2 + \lambda \vert \nabla u \vert ^2 - 2 \gamma_2 u \vert \nabla u \vert ^2 - 3\gamma_3 u^2 \vert \nabla u \vert ^2 - \sigma u^2 \, \mathrm{d} x \\
& \leq \int_\Omega \vert \nabla u \vert ^2 (\lambda + 2 \vert \gamma_2 u \vert - 3 \gamma_3 u^2) - \sigma u^2 \, \mathrm{d}x \\
& \leq \int_\Omega \vert \nabla u \vert ^2 (\lambda -\gamma_3 u^2 - 2\gamma_3 ( \vert u \vert -\fraction{\vert \gamma_2 \vert}{2 \gamma_3})^2 + \fraction{\gamma_2^2}{2\gamma_3}) -\sigma u^2  \, \mathrm{d}x \\
& \leq \int_\Omega \vert \nabla u \vert ^2 (\lambda + \fraction{\gamma_2^2}{2\gamma_3}) -\sigma u^2 \, \mathrm{d}x \\
& < 0.
\end{aligned}
\end{equation*}
Therefore, when $\lambda < \widetilde{\lambda}$, (\ref{Eqn. 2}) has
no non-trivial singular points in $H$. It is known that for any
$\lambda \in \mathbb{R}^1$, (\ref{Eqn. 2}) possesses a global
attractor. Thus condition (i) follows. Condition (ii) is a direct consequence of the
existence of global attractors for any $\lambda \in \mathbb{R}^1$.
The claim for saddle-node bifurcations is thus proven, and the proof
is complete. \ep

\bp[Proof of Theorem \ref{bif. thm hex}] The eigenvectors of
$L_\lambda$ corresponding to $K_1^c$ and $K_2^c$ are
\begin{equation}
\begin{aligned}
& e_1 = \cos(\frac{n}{d}x_1), \\
& e_2 = \cos(\frac{n}{2d}x_1)\cos(\frac{\sqrt{3}n}{2d}x_2).
\end{aligned}
\end{equation}

Let $u=v + \Phi(v, \lambda)$, where $v=y_1e_1 + y_2e_2$ and $\Phi$
is the center manifold function. Again, $\Phi$ can be approximated
to the second order via the formula given in \eqref{abstract cmd
app}. In this case, we have
\begin{equation}
\begin{aligned}
\Phi(y,\lambda) = & -\frac{2\gamma_2|K^c_1|^2 y_1^2}{14|K^c_1|^4 -
2\lambda |K^c_1|^2 -\sigma}\cos(\frac{2n}{d}x_1) \\
& - \frac{3\gamma_2 |K^c_1|^2 y_1y_2}{7|K^c_1|^4 - \lambda
|K^c_1|^2 - \sigma}\cos(\frac{3n}{2d}x_1)\cos(\frac{\sqrt{3}n}{2d}x_2) \\
 &- \frac{\gamma_2 |K^c_1|^2 y_2^2}{14|K^c_1|^4 - 2\lambda
 |K^c_1|^2 - \sigma}\cos(\frac{n}{d}x_1)\cos(\frac{\sqrt{3}n}{d}x_2) \\
 &- \frac{3\gamma_2 |K^c_1|^2 y_2^2}{28|K^c_1|^4 - 4\lambda
|K^c_1|^2 - 4\sigma}\cos(\frac{\sqrt{3}n}{d}x_2) + o(|y|^2),
\end{aligned}
\end{equation}
where $|K_1^c| = |K_2^c|=n/L$.

The reduced equation to the center manifold is now given by
\begin{equation} \label{reduced eqn hex case}
\begin{aligned}
\frac{\mathrm{d} y_1}{\mathrm{d} t}  &=  \beta_1 (\lambda)
y_1 - \frac{|K^c_1|^2 \gamma_2}{4} y_2^2 + by_1^3 + cy_1y_2^2 + o(|y|^3), \\
\frac{\mathrm{d} y_2}{\mathrm{d} t}  &=  \beta_1 (\lambda) y_2 -
|K^c_1|^2 \gamma_2 y_1y_2 + dy_2^3 + ey_1^2y_2 + o(|y|^3),
\end{aligned}
\end{equation}
where $\beta_1(\lambda) = \beta_{K_1^c}(\lambda) =
\beta_{K_2^c}(\lambda)$ and
\begin{equation*}
\begin{aligned}
b&=\frac{2|K^c_1|^4\gamma_2^2}{14|K^c_1|^4 - 2\lambda|K^c_1|^2-\sigma} -
\frac{3|K^c_1|^2}{4}\gamma_3, \\
c&= \frac{3|K^c_1|^4\gamma_2^2}{14|K^c_1|^4 - 2\lambda|K^c_1|^2 - 2\sigma}
- \frac{3|K^c_1|^2}{4}\gamma_3, \\
d &= \frac{|K^c_1|^4\gamma_2^2}{28|K^c_1|^4 - 4\lambda|K^c_1|^2-2\sigma} +
\frac{3|K^c_1|^4\gamma_2^2}{28|K^c_1|^4 - 4\lambda|K^c_1|^2-4\sigma} -
\frac{9|K^c_1|^2}{16}\gamma_3,\\
e&=\frac{3|K^c_1|^4\gamma_2^2}{7|K^c_1|^4 - \lambda|K^c_1|^2 - \sigma} -
\frac{3|K^c_1|^2}{2}\gamma_3.
\end{aligned}
\end{equation*}
As before, we have $|K^c_1|^2 \simeq \sqrt{\sigma}$. Then when $\lambda$ is close to $\lambda_c$, the above coefficients can be written as
\begin{equation} \label{b-e}
\begin{aligned}
b& \simeq \frac{2}{9}\gamma_2^2 -
\frac{3\sqrt{\sigma}}{4}\gamma_3 + o(|\lambda - \lambda_c|), \\
c& \simeq \frac{3}{8}\gamma_2^2 - \frac{3\sqrt{\sigma}}{4}\gamma_3 + o(|\lambda - \lambda_c|), \\
d & \simeq \frac{35}{144}\gamma_2^2  - \frac{9 \sqrt{\sigma}}{16}\gamma_3 + o(|\lambda - \lambda_c|),\\
e& \simeq \frac{3}{4} \gamma_2^2 -
\frac{3\sqrt{\sigma}}{2}\gamma_3 + o(|\lambda - \lambda_c|).
\end{aligned}
\end{equation}
It is known that the transition of \eqref{reduced eqn hex case} near
$\lambda = \lambda_c$ can be understood by studying the behavior of
the equation when $\lambda$ is at the critical point $\lambda_c =
(|K^c_1|^4 + \sigma)/|K^c_1|^2$. At $\lambda = \lambda_c$,
$\beta_1(\lambda)=0$, and \eqref{reduced eqn hex case} reads (up to
3rd order terms):
\begin{equation} \label{reduced eqn hex case critical pt}
\begin{aligned}
\frac{\mathrm{d} y_1}{\mathrm{d} t}  &=  - \frac{|K^c_1|^2 \gamma_2}{4} y_2^2 + by_1^3 + cy_1y_2^2, \\
\frac{\mathrm{d} y_2}{\mathrm{d} t}  &=  - |K^c_1|^2 \gamma_2 y_1y_2 +
dy_2^3 + ey_1^2y_2.
\end{aligned}
\end{equation}

Note that by \eqref{b-e} we have $b + 4c = 4d + e$, from which we
see that on the straight lines $y_2 = \pm 2y_1$, the equation
\eqref{reduced eqn hex case critical pt} satisfies that
\begin{equation*}
\frac{\mathrm{d} y_2}{\mathrm{d} y_1} =\pm2, \text{ for all } y_1
\text{ such that } -|K^c_1|^2 \gamma_2y_1^2 + (b+ 4c)y_1^3 \neq 0.
\end{equation*}
Hence, the straight lines $y_2 = \pm 2y_1$ are orbits of \eqref{reduced eqn hex case critical pt}. Obviously, $y_2=0$ also consists of orbits of \eqref{reduced eqn hex case critical pt}. If $\gamma_2\neq 0$, all straight line orbits of \eqref{reduced eqn hex case critical pt} lie on one of these three lines. If $\gamma_2=0$, $y_1=0$ also consists of orbits of \eqref{reduced eqn hex case critical pt}, and in this case all straight line orbits lie on one of four straight lines $y_2=\pm 2y_1$, $y_2 = 0$ or $y_1=0$.

\begin{figure}
    \subfigure[]{
    \label{fig a=0 beta=0}
    \begin{minipage}[b]{0.45\textwidth}
    \centering
   \includegraphics [height=0.6\textwidth, width=0.6\textwidth, trim = 0 0 0 0, clip]{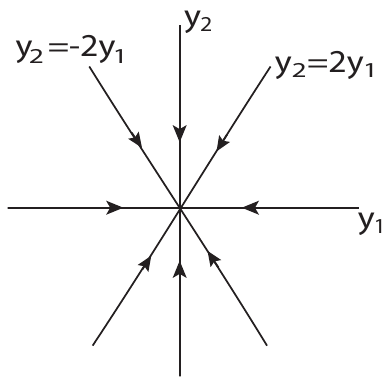}
    \end{minipage}}
    \subfigure[]{
    \label{fig a=0 beta>0}
    \begin{minipage}[b]{0.45\textwidth}
    \centering
   \includegraphics [height=0.6\textwidth, width=0.6\textwidth, trim = 0 0 0 0, clip]{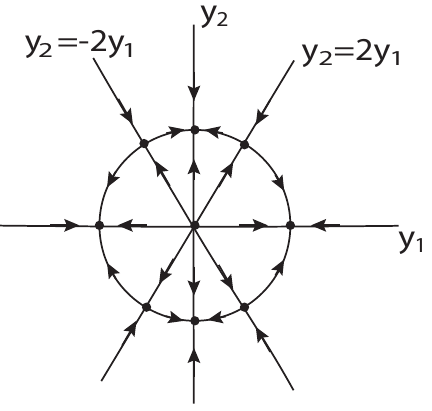}
    \end{minipage}}
  \caption{The topological structure of flows of \eqref{reduced eqn hex case}, (a) for $\gamma_2=0$ and $\beta_1(\lambda)=0$, and (b) for $\gamma_2=0$ and $\beta_1(\lambda)>0$.}
\end{figure}

For the case $\gamma_2 = 0$, the numbers $b$, $c$, $d$, and $e$ are all less
than zero by \eqref{gamma_3} and \eqref{b-e}. It is easy to see that all the straight
line orbits of \eqref{reduced eqn hex case critical pt} tend to the
origin as shown in Figure \ref{fig a=0 beta=0}, which implies that
the origin $(y_1, y_2) = (0,0)$ is locally asymptotically stable.
Therefore, the transition of \eqref{reduced eqn hex case} is Type-I
\cite{MW05}. Omitting $o(|y|^3)$ terms in \eqref{reduced eqn hex case},
the bifurcated structure are shown in Figure \ref{fig a=0 beta>0}.
The origin becomes unstable, and there are eight other steady states
emerging from the transition, among which four of them are stable and
the rest are saddle points. The stability of the steady states can
be analyzed by linearizing \eqref{reduced eqn hex case} about the
corresponding steady state, and the results are shown in Figure
\ref{fig a=0 beta>0}. By the center manifold reduction, the
bifurcated structure for the original equation \eqref{Eqn. 2} is a
perturbation of the structure shown in Figure \ref{fig a=0 beta>0},
and Assertion (1) is thus proved.
\begin{figure}
    \subfigure[]{
    \label{fig a>0 b>0}
    \begin{minipage}[b]{0.45\textwidth}
    \centering
   \includegraphics [height=0.6\textwidth, width=0.6\textwidth, trim = 0 0 0 0, clip]{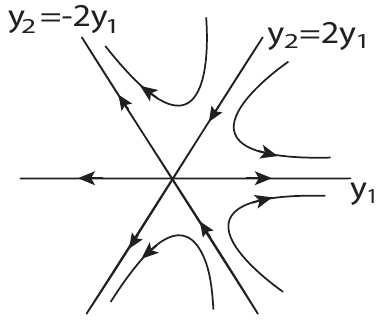}
    \end{minipage}}
   \subfigure[]{
    \label{fig a<0 b>0}
    \begin{minipage}[b]{0.45\textwidth}
    \centering
   \includegraphics [height=0.6\textwidth, width=0.6\textwidth, trim = 0 0 0 0, clip]{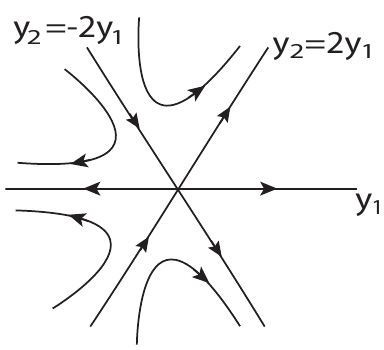}
    \end{minipage}}
\subfigure[]{
    \label{fig a>0 b<0}
    \begin{minipage}[b]{0.45\textwidth}
    \centering
   \includegraphics [height=0.6\textwidth, width=0.6\textwidth, trim = 0 0 0 0, clip]{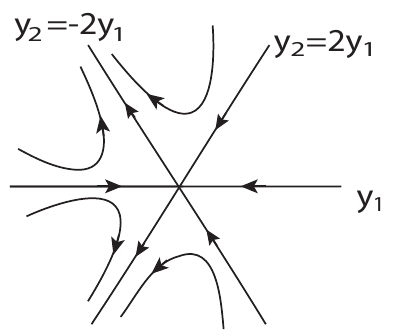}
    \end{minipage}}
    \subfigure[]{
    \label{fig a<0 b<0}
    \begin{minipage}[b]{0.45\textwidth}
    \centering
   \includegraphics [height=0.6\textwidth, width=0.6\textwidth, trim = 0 0 0 0, clip]{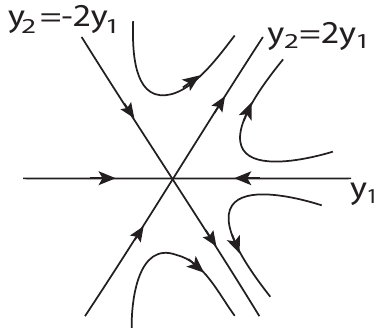}
    \end{minipage}}
 \caption{The topological structure of flows of \eqref{reduced eqn hex case critical pt}, (a) for $\gamma_2>0$ and $b>0$, (b) for $\gamma_2<0$ and $b>0$, (c) for $\gamma_2>0$ and $b<0$, and (b) for $\gamma_2<0$ and $b<0$.}
\end{figure}

When $\gamma_2\neq 0$, the straight line orbits of \eqref{reduced
eqn hex case critical pt} consist of $y_2 = \pm 2y_1$ and $y_2=0$ as
shown before. The phase diagram of \eqref{reduced eqn hex case
critical pt} near the origin can now be easily determined and are
shown in Figures \ref{fig a>0 b>0} -- \ref{fig a<0 b<0}.

For $\gamma_2 \neq 0$ and $b>0$,
a neighborhood of the origin is divided into six regions by the
straight line orbits, and four of them are hyperbolic and the rest two
regions are parabolic as shown in Figures \ref{fig a>0 b>0} and
\ref{fig a<0 b>0}. Since orbits starting from any of these regions
are eventually repelled away from the origin, the transition is
Type-II \cite{MW}. The number and type of equilibrium points
bifurcated out for each case can be analyzed in the same fashion as
in the proof for Assertion (1), thus Assertion (2) is proved.

\begin{figure}
\subfigure[]{
    \label{fig type III flow a>0}
    \begin{minipage}[b]{0.45\textwidth}
    \centering
   \includegraphics [height=0.6\textwidth, width=0.6\textwidth, trim = 0 0 0 0, clip]{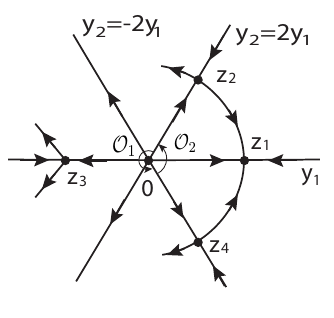}
    \end{minipage}}
    \subfigure[]{
    \label{fig type III flow a<0}
    \begin{minipage}[b]{0.45\textwidth}
    \centering
   \includegraphics [height=0.6\textwidth, width=0.6\textwidth, trim = 0 0 0 0, clip]{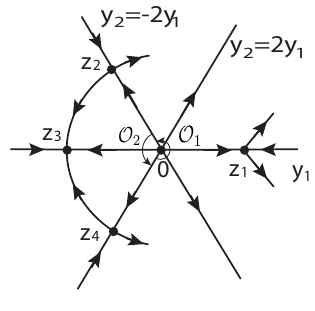}
    \end{minipage}}
 \caption{The topological structure of flows of \eqref{reduced eqn hex case} for $\lambda > \lambda_c$, $b<0$ and (a) for $\gamma_2>0$, (b) for $\gamma_2<0$. The two sectorial regions $\mathcal{O}_1$ and $\mathcal{O}_2$ are seperated by the rays $\overline{0z_2}$ and $\overline{0z_4}$.}
 \label{type III flow}

\end{figure}

For $\gamma_2 \neq 0$ and $b<0$, a neighborhood of the origin is also
divided into six regions by the straight line orbits, and four of
them are hyperbolic and the rest two regions are parabolic as shown
in Figures \ref{fig a>0 b<0} and \ref{fig a<0 b<0}. Here, orbits in
the parabolic regions tend to the origin, and the transition is
Type-III \cite{MW}. The flow structure of the reduced equation \eqref{reduced eqn hex case} when $\lambda > \lambda_c$ can also be obtained easily and the result after dropping $o(|y|^3)$ terms is shown in Figure \ref{type III flow}.
From here, we see that there is a neighborhood $\mathcal{N}$ of the origin of the $y_1y_2$ plane, which can be decomposed into two
disjoint regions $\mathcal{N}_1$ and $\mathcal{N}_2$ such that $\overline{\mathcal{N}}=
\overline{\mathcal{N}}_1 \cup \overline{\mathcal{N}}_2$ and in region $\mathcal{N}_2$, there is exactly one stable node bifurcating out on the side $\lambda>\lambda_c$. Actually, $\mathcal{N}_1$ and $\mathcal{N}_2$ are slight perturbations of $\mathcal{O}_1$ and $\mathcal{O}_2$ respectively as shown in Figure \ref{type III flow}. Corresponding to $\mathcal{N}$, there is a neigborhood $V$ of the origin in space $H$, which can be decomposed into two
disjoint regions $V_1$ and $V_2$ such that $\overline{V}= \overline{V_1} \cup \overline{V_2}$. Here $V_1$ and $V_2$ are chosen such that
\begin{equation} \label{V}
(1-P_2) V_i = \mathcal{N}_i, \: 1\le i \le 2,
\end{equation}
where $P_2$ is the canonical projection as in \eqref{abstract cmd app}. Thus, assertion (3) is proved and the proof is complete.
\ep

The proofs of Theorem \ref{bif. thm case 1.2} and Theorem \ref{bif. thm case 1.3} are similar to that of Theorem
\ref{bif. thm}, and the details are omitted here. But for the
convenience of the readers, we present the center manifold functions
and the reduced equations to the center manifold.

\bp[Proof of Theorem \ref{bif. thm case 1.2}] By similar
calculation, we have the following center manifold function:
\begin{eqnarray*} 
\Phi(y,\lambda)  = -\xi_1 \gamma_2 y^2 \cos(\fraction{2k_1\pi x_1}{L_1}) -\xi_2 \gamma_2 y^2 \cos(\fraction{2k_2\pi x_2}{L_2})  \nonumber \\
- \xi_3 \gamma_2 y^2 \cos(\fraction{2k_1\pi
x_1}{L_1})\cos(\fraction{2k_2\pi x_2}{L_2}) + o(\vert y \vert
^2).
\end{eqnarray*}
Here $\xi_1$, $\xi_2$ and $\xi_3$ are as follows:
\begin{align*} 
& \xi_1(\lambda) = \fraction{\vert K_{11} \vert^2}{14 \vert K_{11} \vert^4  -4 \vert K_{11} \vert^2 \vert K_{12} \vert^2  - 2 \vert K_{12} \vert^4 + 2 \lambda (\vert K_{12} \vert^2 - \vert K_{11} \vert^2) - \sigma},  \\ 
&\xi_2(\lambda) = \fraction{\vert K_{12} \vert^2}{14 \vert K_{12} \vert^4 -4 \vert K_{11} \vert^2\vert K_{12} \vert^2 - 2 \vert K_{11} \vert^4 + 2 \lambda (\vert K_{11} \vert^2 - \vert K_{12} \vert^2) - \sigma}, \\ 
&\xi_3(\lambda) = \fraction{\vert K_1 \vert^2}{14 \vert K_1 \vert^4 - 2 \lambda \vert K_1 \vert^2 - \sigma},  
\end{align*}
where $\vert K_{11} \vert^2 = \fraction{k_1^2 \pi^2}{L_1^2}, \quad \vert K_{12} \vert^2 = \fraction{k_2^2 \pi^2}{L_2^2}, \quad \vert K_1 \vert^2 = \fraction{k_1^2 \pi^2}{L_1^2} + \fraction{k_2^2 \pi^2}{L_2^2}.$ 

The reduced equations of (\ref{Eqn. 2}) to the center manifold in this case is:
\begin{equation*} 
\frac{\mathrm{d} y}{\mathrm{d} t}   =  \beta_{K_1} (\lambda) y -
\vert K_1 \vert^2 \Big (\fraction{9}{16}\gamma_3 -
\big(\xi_1(\lambda) + \xi_2(\lambda) + \fraction{1}{2}
\xi_3(\lambda)\big)\gamma_2^2\Big )y^3 + o(\vert y \vert ^3).
\end{equation*}

\ep

\bp[Proof of Theorem \ref{bif. thm case 1.3}] The center manifold
function for this case is:
\begin{align*} 
\Phi(y,\lambda) &= -\eta_1 \gamma_2 y^2 \cos(\fraction{2k_1\pi x_1}{L_1}) -\eta_2 \gamma_2 y^2 \cos(\fraction{2k_2\pi x_2}{L_2})  \nonumber \\
&- \eta_3 \gamma_2 y^2 \cos(\fraction{2k_3\pi x_3}{L_3}) - \eta_4
\gamma_2 y^2 \cos(\fraction{2k_1\pi
x_1}{L_1}) \cos(\fraction{2k_2\pi x_2}{L_2}) \\
& - \eta_5 \gamma_2 y^2 \cos(\fraction{2k_1\pi x_1}{L_1})
\cos(\fraction{2k_3\pi x_3}{L_3})- \eta_6 \gamma_2 y^2
\cos(\fraction{2k_2\pi
x_2}{L_2}) \cos(\fraction{2k_3\pi x_3}{L_3}) \nonumber \\
& - \eta_7 \gamma_2 y^2 \cos(\fraction{2k_1\pi x_1}{L_1})
\cos(\fraction{2k_2\pi x_2}{L_2}) \cos(\fraction{2k_3\pi
x_3}{L_3}) + o(\vert y \vert ^2), \nonumber
\end{align*}

and the reduced equations of (\ref{Eqn. 2}) to the center manifold is given by:
\begin{eqnarray*} 
\frac{\mathrm{d} y}{\mathrm{d} t}   =  \beta_{K_1} (\lambda) y -
\vert K_1 \vert^2 (\fraction{27}{64}\gamma_3 - \eta \gamma_2^2)y^3 +
o(\vert y \vert ^3),
\end{eqnarray*}
Here $\eta$ and $\eta_i$ $(1\leq i \leq 7)$ are given by 
\begin{equation*}
\begin{aligned}
& \eta_1(\lambda) =\fraction{k_1^2 \pi^2 / L_1^2}{2\bigl [-2\vert K_{1} \vert^4  +\fraction{16 k_1^4\pi ^4}{L_1^4} + \lambda (2\vert K_{1} \vert^2 - \fraction{4 k_1^2\pi ^2}{L_1^2}) - \sigma \bigr]}, \\
& \eta_2(\lambda) =\fraction{k_2^2 \pi^2 / L_2^2}{2\bigl[-2\vert K_{1} \vert^4  +\fraction{16 k_2^4\pi ^4}{L_2^4} + \lambda (2\vert K_{1} \vert^2 - \fraction{4 k_2^2\pi ^2}{L_2^2}) - \sigma\bigr]},  \\
& \eta_3(\lambda) =\fraction{k_3^2 \pi^2 / L_3^2}{2\bigl[-2\vert K_{1} \vert^4  +\fraction{16 k_3^4\pi ^4}{L_3^4} + \lambda (2\vert K_{1} \vert^2 - \fraction{4 k_3^2\pi ^2}{L_3^2}) - \sigma\bigr]}, \\ 
& \eta_4(\lambda) = \fraction{\vert K_{12} \vert ^2}{2\bigl[-2\vert K_{1} \vert^4  +16\vert K_{12} \vert ^4 + 2\lambda (\vert K_{1} \vert^2 - 2\vert K_{12} \vert ^2) - \sigma\bigr]},\\  
& \eta_5(\lambda) = \fraction{\vert K_{13} \vert ^2}{2\bigl[-2\vert K_{1} \vert^4  +16\vert K_{13} \vert ^4 + 2\lambda (\vert K_{1} \vert^2 - 2\vert K_{13} \vert ^2) - \sigma\bigr]}, \\
& \eta_6(\lambda) = \fraction{\vert K_{23} \vert ^2}{2\bigl[-2\vert K_{1} \vert^4  +16\vert K_{23} \vert ^4 + 2\lambda (\vert K_{1} \vert^2 - 2\vert K_{23} \vert ^2) - \sigma\bigr]}, \\ 
& \eta_7(\lambda) = \fraction{\vert K_{1} \vert ^2}{14\vert K_{1} \vert^4 - 2\lambda \vert K_{1} \vert^2  - \sigma}, \\
&\eta(\lambda) = \eta_1 + \eta_2 + \eta_3 + \fraction{1}{2} (\eta_4 + \eta_5 + \eta_6) + \fraction{1}{4} \eta_7,
\end{aligned}
\end{equation*}
\noindent where
\begin{align*}
&\vert K_{12} \vert^2 = \fraction{k_1^2\pi ^2}{L_1^2} + \fraction{k_2^2\pi ^2}{L_2^2}, & &\vert K_{13} \vert^2 = \fraction{k_1^2\pi ^2}{L_1^2} + \fraction{k_3^2\pi ^2}{L_3^2}, \\ &\vert K_{23} \vert^2 = \fraction{k_2^2\pi ^2}{L_2^2} + \fraction{k_3^2\pi ^2}{L_3^2}, & &\vert K_1 \vert^2 = \fraction{k_1^2 \pi^2}{L_1^2} + \fraction{k_2^2 \pi^2}{L_2^2} +  \fraction{k_3^2 \pi^2}{L_3^2}.
\end{align*} 

\ep

\bibliographystyle{plain}

\bibliography{Bibl}

\begin{thebibliography}{10}

\bibitem{BF90}
F.~S. Bates and G.~H. Fredrickson.
\newblock Block copolymer thermodynamics: theory and experiment.
\newblock {\em Annu. Rev. Phys. Chem.}, 41:525--557, 1990.

\bibitem{BF99}
F.~S. Bates and G.~H. Fredrickson.
\newblock Block copolymers -- designer soft materials.
\newblock {\em Physics Today}, 52:32--38, 1999.

\bibitem{BM96}
F.~S. Bates and M.~W. Matsen.
\newblock Unifying weak- and strong-segragation block copolymer theories.
\newblock {\em Macromolecules}, 29(4):1091--1098, 1996.

\bibitem{Cah68}
J.~W. Cahn.
\newblock Spinodal decomposition.
\newblock {\em Tran. Meta. Soc. AIME}, 242:166--180, 1968.

\bibitem{CPW09}
R.~Choksi, M.~A. Peletier, and J.~F. Williams.
\newblock On the phase diagram for microphase separation of diblock copolymers:
  An approach via a nonlocal ~cahn-hilliard functional.
\newblock {\em SIAM J. Appl. Math.}, 69(6):1712--1738, 2009.

\bibitem{CR03}
R.~Choksi and X.~Ren.
\newblock On the derivation of a density functional theory for microphase
  separation of diblock copolymers.
\newblock {\em J. Stat. Phys.}, 113:151--176, 2003.

\bibitem{deG79}
P.~de~Gennes.
\newblock {\em Scaling concepts in polymer physics}.
\newblock Cornell university press, 1979.

\bibitem{deG80}
P.~de~Gennes.
\newblock Dynamics of fluctuations and spinodal decomposition in polymer
  blends.
\newblock {\em J. Chem. Phys.}, 72(1):4756--4763, 1980.

\bibitem{DK09}
R.~C. Desai and R.~Kapral.
\newblock {\em Dynamics of Self-Organized and Self-Assembled Structures}.
\newblock Cambridge University Press, 2009.

\bibitem{FK51}
P.~J. Flory and W.~R. Krigbaum.
\newblock Thermodynamics of high polymer solutions.
\newblock {\em Annu. Rev. Phys. Chem.}, 2:383--402, 1951.

\bibitem{FVD02}
G.~H. Fredrickson, V.~Ganesan, and F.~Drolet.
\newblock Field-theoretic computer simulation methods for polymers and complex
  fluids.
\newblock {\em Macromolecules}, 35:16--39, 2002.

\bibitem{Hen81}
D.~Henry.
\newblock {\em Geometric theory of semilinear parabolic equations}, volume 840
  of {\em Lecture Notes in Mathematics}.
\newblock Springer-Verlag, Berlin, 1981.

\bibitem{Hug41}
M.~L. Huggins.
\newblock Solutions of long chain compounds.
\newblock {\em J. Chem. Phys.}, 9:440, 1941.

\bibitem{JWHZ}
K.~Jiang, C.~Wang, Y.~Huang, and P.~Zhang.
\newblock Metastable patterns in diblock copolymers.
\newblock {\em submitted to J. chem. Phys.}

\bibitem{Kaw04}
T.~Kawakatsu.
\newblock {\em Statistical physics of polymers}.
\newblock Springer, 2004.

\bibitem{KFBHRBAM95}
A.~K. Khandpur, S.~F\"oster, F.~S. Bates, I.~W. Hamley, A.~J. Ryan, W.~Bras,
  K.~Almdal, and K.~Mortensen.
\newblock Polyisoprene-polystyrene diblock copolymer phase diagram near the
  order-disorder transition.
\newblock {\em Macromolecules}, 28(26):8796--8806, 1995.

\bibitem{MW}
T.~Ma and S.~Wang.
\newblock {\em Phase Transition Dynamics in Nonlinear Sciences}.
\newblock to appear.

\bibitem{MW05}
T.~Ma and S.~Wang.
\newblock {\em Bifurcation theory and applications}, volume~53 of {\em World
  Scientific Series on Nonlinear Science. Series A: Monographs and Treatises}.
\newblock World Scientific Publishing Co. Pte. Ltd., Hackensack, NJ, 2005.

\bibitem{MW09a}
T.~Ma and S.~Wang.
\newblock Cahn hilliard equations and phase transition dynamics for binary
  systems.
\newblock {\em Disc. Cont. Dyn. Sys. B}, 11:741--784, 2009.

\bibitem{MS94}
M.~W. Matsen and M.~Schick.
\newblock Stable and unstable phases of a diblock copolymer melt.
\newblock {\em Phys. Rev. lett}, 72(16):2660--2663, 1994.

\bibitem{NO95}
Y.~Nishiura and I.~Ohnishi.
\newblock Some mathematical aspects of the ~micro-phase separation in diblock
  copolymers.
\newblock {\em Physica D}, 84:31--39, 1995.

\bibitem{NS84}
A.~Novick-Cohen and L.~A. Segel.
\newblock Nonlinear aspects of the cahn-hilliard equation.
\newblock {\em Physica D}, 10:277--298, 1984.

\bibitem{OK86}
T.~Ohta and K.~Kawasaki.
\newblock Equilibrium morphology of blcok copolymer melts.
\newblock {\em Macromolecules}, 19(10):2621--2632, 1986.

\bibitem{OE97}
F.~Otto and W.~E.
\newblock Thermodynamically driven incompressible fluid mixtures.
\newblock {\em J. Chem. Phys.}, 107:10177--10184, 1997.

\bibitem{Rei98}
L.~E. Reichl.
\newblock {\em A modern course in statistical physics}.
\newblock Wiley-Interscience, New York, 1998.

\bibitem{SA95}
M.~Seul and D.~Andelman.
\newblock Domain shapes and patterns: the phenomenology of modulated phases.
\newblock {\em Science}, 267:476--483, 1995.

\bibitem{Str07}
G.~Strobl.
\newblock {\em The Physics of Polymers -- Concepts for Understanding Their
  Structures and Behavior}.
\newblock Springer, 2007.

\bibitem{Tem97}
R.~Temam.
\newblock {\em Infinite-dimensional dynamical systems in mechanics and
  physics}, volume~68 of {\em Applied Mathematical Sciences, second edition}.
\newblock Springer-Verlag, New York, 1997.

\end{thebibliography}

\end{document}